\documentclass[twocolumn]{aastex631}

\newcommand{\s}{\ensuremath{\mbox{~s}}}

\newcommand{\cm}{\ensuremath{\mbox{~cm}}}
\newcommand{\pcmsq}{\ensuremath{\cm^{-2}}}

\newcommand{\Hii}{H\textsc{ii}}

\newcommand{\vel}{km\,s$^{-1}$}

\newcommand{\msun}{$M_{\odot}$}
\newcommand{\lsun}{$L_{\odot}$}
\newcommand{\um}{$\mu$m}

\newcommand{\egcite}{\citep[e.g.,][]}

\newcommand{\hcop}{HCO$^{+}$}
\newcommand{\htcop}{H$^{13}$CO$^{+}$}

\newcommand{\uchii}{UC-H\textsc{ii}}

\newcommand{\mdotyr}{$M_{\odot}$~yr$^{-1}$}

\usepackage{ae,aecompl}
\usepackage{subfigure}
\usepackage{graphicx}	
\usepackage{amsmath}	
\usepackage{amssymb}	
\usepackage{color,xcolor}
\usepackage{float}

\def   \aj {{\rm {AJ}}}

\def   \apj {{\rm {ApJ}}}

\def   \apjs {{\rm {ApJS}}}
\def   \apjl {{\rm {ApJL}}}

\def   \aap {{\rm {A\&A}}}

\def   \aaps {{\rm {A\&AS}}}

\def   \mnras {{\rm {MNRAS}}}
\def   \pasp {{\rm {PASP}}}

\graphicspath{{./G310.1420+0.7583/}{figures/}}

\received{xxx}
\revised{xxx}
\accepted{xxx}

\shorttitle{Multi-scale mass accretion in high-mass star formation}
\shortauthors{Yang et al.}

\begin{document}
\title{\bf Direct observational evidence of the multi-scale, dynamical mass accretion toward a high-mass star forming hub-filament system}

\submitjournal{ApJ}
\correspondingauthor{Hong-Li Liu}
\email{hongliliu2012@gmail.com}

\author{Dongting Yang}
\affiliation{School of physics and astronomy, Yunnan University, Kunming, 650091, PR China}

\author{Hong-Li Liu}
\affiliation{School of physics and astronomy, Yunnan University, Kunming, 650091, PR China}

\author{Anandmayee Tej}
\affiliation{Indian Institute of Space Science and Technology, Thiruvananthapuram 695 547, Kerala, India}
\correspondingauthor{Anandmayee Tej}
\email{tej@iist.ac.in}

\author{Tie Liu}
\affiliation{Shanghai Astronomical Observatory, Chinese Academy of Sciences, 80 Nandan Road, Shanghai 200030, Peoples Republic of China}
\affiliation{Key Laboratory for Research in Galaxies and Cosmology, Shanghai Astronomical Observatory,Chinese Academy of Sciences, 80 Nandan Road, Shanghai 200030, Peoples Republic of China}

\author{Patricio Sanhueza}
\affiliation{National Astronomical Observatory of Japan, National Institutes of Natural Sciences, 2-21-1 Osawa, Mitaka, Tokyo 181-8588, Japan}
\affiliation{Department of Astronomical Science, The Graduate University for Advanced Studies, SOKENDAI, 2-21-1 Osawa, Mitaka, Tokyo 181-8588, Japan}

\author{Sheng-Li Qin}
\affiliation{School of physics and astronomy, Yunnan University, Kunming, 650091, PR China}

\author{Xing Lu}
\affiliation{Shanghai Astronomical Observatory, Chinese Academy of Sciences, 80 Nandan Road, Shanghai 200030, Peoples Republic of China}

\author[0000-0002-7237-3856]{Ke Wang}
\affiliation{Kavli Institute for Astronomy and Astrophysics, Peking University, 5 Yiheyuan Road, Haidian District, Beijing 100871, People's Republic of China}
\affiliation{Department of Astronomy, Peking University, 100871, Beijing, People's Republic of China}

\author{Sirong Pan}
\affiliation{School of physics and astronomy, Yunnan University, Kunming, 650091, PR China}

\author{Feng-Wei Xu}
\affiliation{Kavli Institute for Astronomy and Astrophysics, Peking University, 5 Yiheyuan Road, Haidian District, Beijing 100871, People's Republic of China}
\affiliation{Department of Astronomy, Peking University, 100871, Beijing, People's Republic of China}

\author[0000-0002-1424-3543]{Enrique Vazquez-Semadeni}
\affiliation{Instituto de Radioastronomía y Astrofísica, Universidad Nacional Autónoma de México, Antigua Carretera a Pátzcuaro 8701, ExHda. San José de la Huerta, Morelia, Michoacán, México C.P. 58089}

\author{Shanghuo Li}
\affiliation{Max Planck Institute for Astronomy, Königstuhl 17, D-69117 Heidelberg, Germany}

 \author[0000-0003-4714-0636]{Gilberto C. Gómez}
\affiliation{Instituto de Radioastronomía y Astrofísica, Universidad Nacional Autónoma de México, Antigua Carretera a Pátzcuaro  8701, ExHda. San José de la Huerta, Morelia, Michoacán, México C.P. 58089}

\author[0000-0002-9569-9234]{Aina Palau}
\affiliation{Instituto de Radioastronomía y Astrofísica, Universidad Nacional Autónoma de México, Antigua Carretera a Pátzcuaro 8701, ExHda. San José de la Huerta, Morelia, Michoacán, México C.P. 58089}

\author{Guido Garay}
\affiliation{Departamento de Astronom\'ia, Universidad de Chile, Casilla 36-D, Santiago, Chile}

\author{Paul F. Goldsmith}
\affiliation{Jet Propulsion Laboratory, California Institute of Technology, 4800 Oak Grove Drive, Pasadena, CA 91109, USA}

\author{Mika Juvela}
\affiliation{Department of Physics, PO box 64, FI- 00014, University of Helsinki,Finland}

\author{Anindya Saha}
\affiliation{Indian Institute of Space Science and Technology, Thiruvananthapuram 695 547, Kerala, India}

\author{Leonardo Bronfman}
\affiliation{Departamento de Astronom\'ia, Universidad de Chile, Casilla 36-D, Santiago, Chile}

\author{Chang Won Lee}
\affiliation{Korea Astronomy and Space Science Institute, 776 Daedeokdaero, Yuseonggu, Daejeon 34055, Republic of Korea}
\affiliation{University of Science and Technology, Korea (UST), 217 Gajeong-ro,
Yuseong-gu, Daejeon 34113, Republic of Korea}

\author{Ken'ichi Tatematsu}
\affiliation{National Astronomical Observatory of Japan, National Institutes of Natural Sciences, 2-21-1 Osawa, Mitaka, Tokyo 181-8588, Japan}

\author{Lokesh Dewangan}
\affiliation{Physical Research Laboratory, Navrangpura, Ahmedabad—380 009, India}

\author{Jianwen Zhou}
\affiliation{Max Planck Institute for Astronomy, Königstuhl 17, D-69117 Heidelberg, Germany}

\author{Yong Zhang}
\affiliation{School of Physics and Astronomy, Sun Yat-sen University, 2 Daxue Road, Zhuhai, Guangdong, 519082, People's Republic of China}

\author{Amelia Stutz}
\affiliation{Departamento de Astronom\'ia, Universidad de Concepci\'on, Av. Esteban Iturra s/n, Distrito Universitario, 160-C, Chile}
\affiliation{Max Planck Institute for Astronomy, Königstuhl 17, D-69117 Heidelberg, Germany}

\author{Chakali Eswaraiah}
\affiliation{Indian Institute of Science Education and Research Tirupati, Rami Reddy Nagar, Karakambadi Road, Mangalam (P.O.), Tirupati 517 507, India}

\author{L. Viktor Toth}
\affiliation{E\"{o}tv\"{o}s Lor\'{a}nd University, Department of Astronomy, P\'{a}zm\'{a}ny P\'{e}ter s\'{e}t\'{a}ny 1/A, H-1117, Budapest, Hungary}

\author{Isabelle Ristorcelli}
\affiliation{IRAP, Universit´e de Toulouse, CNRS, UPS, CNES, Toulouse, France}

\author{Xianjin Shen}
\affiliation{School of physics and astronomy, Yunnan University, Kunming, 650091, PR China}

\author{Anxu Luo}
\affiliation{School of physics and astronomy, Yunnan University, Kunming, 650091, PR China}

\author{James O. Chibueze}
\affiliation{Centre for Space Research, North-West University, Potchefstroom 2520, South Africa}
\affiliation{Department of Mathematical Sciences, University of South Africa, Cnr Christian de Wet Rd and Pioneer Avenue, Florida Park, 1709, Roodepoort, South Africa}
\affiliation{Department of Physics and Astronomy, Faculty of Physical Sciences,  University of Nigeria, Carver Building, 1 University Road,Nsukka 410001, Nigeria}

\begin{abstract}
There is growing evidence that high-mass star formation and hub-filament systems (HFS) are intricately linked. The gas kinematics along the filaments and the forming high-mass star(s) in the central hub are in excellent agreement with the new generation of global hierarchical high-mass star formation models. In this paper, we present an observational investigation of a typical HFS cloud, G310.142+0.758 (G310 hereafter) which reveals unambiguous evidence of mass inflow from the cloud scale via the filaments onto the forming protostar(s) at the hub conforming with the model predictions. Continuum and molecular line data from the ATOMS and MALT90 surveys are used that cover different spatial scales. Three filaments (with total mass $5.7\pm1.1\times 10^3$\,\msun) are identified converging toward the central hub region where several signposts of high-mass star formation have been observed. The hub region contains a massive clump ($1280\pm260$\,\msun) harbouring a central massive core. Additionally, five outflow lobes are associated with the central massive core implying a forming cluster. The observed large-scale, smooth and coherent velocity gradients from the cloud down to the core scale, and the signatures of infall motion seen in the central massive clump and  core, clearly unveil a nearly-continuous, multi-scale mass accretion/transfer process at a similar mass infall rate of $\sim 10^{-3}$\,\mdotyr\ over all scales, feeding the central forming high-mass protostar(s) in the G310 HFS cloud.

\end{abstract}


\section{Introduction} \label{sec:intro}
Filamentary molecular clouds are not only observed to be ubiquitous in the interstellar medium (ISM), but are intimately related to star formation \egcite{And10,Mol10}. Of crucial importance are the special web of filaments known as the hub-filament systems (HFS) which have opened a whole new window to probe the initial stages of high-mass star formation \egcite{Mye09,Kum20,Liu23}. The HFS comprises of at least three filaments apparently converging toward the central web node which is defined as the hub while the associated individual filaments as the hub-composing filaments \egcite{Mye09,Kum20,Liu23}. From the theoretical perspective, the latest generation of star formation models, such as ``global hierarchical collapse'' (GHC, \citealt{V19}) and ``inertial-inflow'' (I2, \citealt{Pad20}), 
advocate the multi-scale mass accretion scenario from clouds to the seeds of star formation. Here, HFSs work as ``conveyor belts" to transport gas material from large-scale clouds, through
hub-composing filaments, to central hubs and further to smaller scales  (i.e., cores, and seeds of star formation). Although this scenario has been strongly supported by recent multi-scale observational and statistical studies \egcite{Kum20,Zho22,Liu23},  GHC advocates a cloud-scale gravitational origin of the flows assembling the HFS, while I2 advocates a turbulent origin. For this reason, the kinetic energy is expected to dominate at the clump scale in the I2 model, while gravitational energy should dominate up to the cloud scale according to GHC.

\begin{figure}
    \centering
    \includegraphics[angle=0, width=0.5\textwidth]{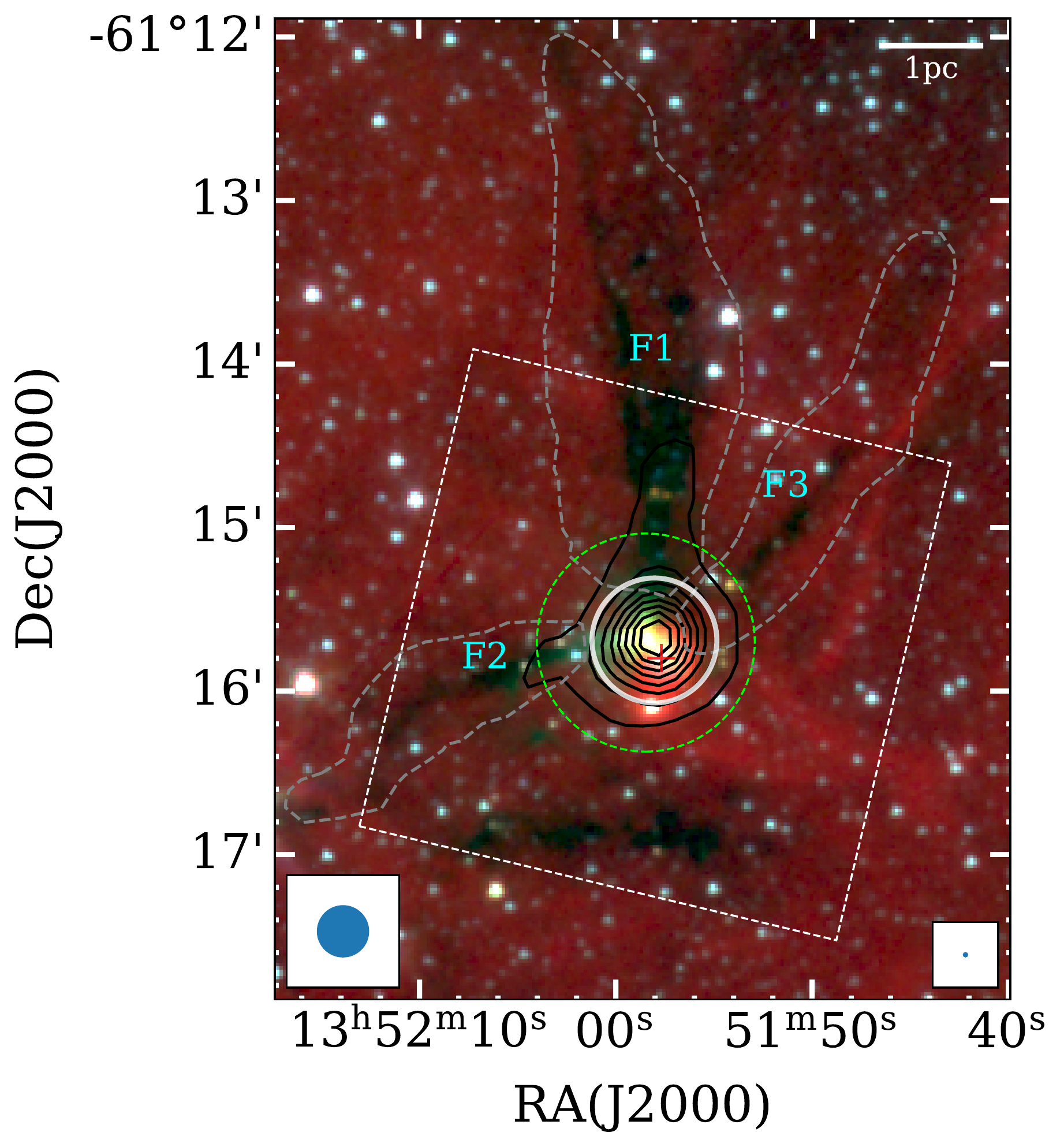} 
    \caption{Three-color composite image of {\it Spitzer} 8.0\,\um\ (red), 4.5\,\um\ (green) and 3.6\,\um\ (blue) overlaid with ATLASGAL 870 {\um} continuum emission (in contours). The three filaments seen as dark dust lanes against bright background emission are labelled F1, F2, and F3.  The contour levels start at 3\,rms with the steps given by the power-law $D=3\times N^{p}+2$, where the dynamical range ($D$) of the intensity map is defined as the ratio of the peak intensity to the rms,  $N$ is the number of the contour levels ($N=[1,8]$ in this case), and the index $p$ is derived from the maximum $N$ and $D$. The red cross shows the location of the \uchii\ region \citep{Pur18}. The white circle shows the extent of the central massive 870\,\um\ clump.The white box and the green circle represent the field view of the MALT90 and ATOMS data, respectively. The polygons (grey dashed line) roughly outline the filamentary structures (F1, F2, F3) based on the $N({\rm H2})$ column density map. The beam sizes of 8.0\,\um\ and 870\,\um\ are displayed at the bottom right and bottom left corners, respectively.}
        
    \label{fig:overview}
\end{figure}

This study is focused toward the HFS cloud G310.142+0.758 (hereafter G310).
It is selected from \citet{Liu23} owing to its distinct HFS structure visible in 
{\it Spitzer} 8.0\,\um\ emission (see Fig.\,\ref{fig:overview}), where a network of three filaments (F1--F3) appear as dark lanes converging towards the cloud centre (defined as the peak of 870\,\um\ emission as shown in contours in the figure). 
A dark east-west filament is visible to the south and does not reveal any apparent connecting feature to the central hub. Hence, we do not consider this filament in our analysis of the HFS.
In addition, the F1 filament seems to contain subfilaments, but are treated here as one entity (see Sect\,\ref{subsec:result:inflow}).
Several distance estimates are available in literature for G310. \cite{Liu21} estimated the distance to be $6.9^{+0.5}_{-0.7}$\,kpc. These authors used the \cite{Rei14} rotation curve with the updated solar motion parameters \citep{Rei19} and the Monte Carlo technique with 10,000 samplings \citep{Wen18}. In this paper, we adopt the distance estimate of $5.4\pm0.5$\,kpc used by many authors \egcite{Bro96,Pur18,Liu20a}. This estimate is 
based on the \cite{Bra93} Galactic rotation curve. Following \citet{Urq14,San17}, we allow for $\sim 10$ percent uncertainty.
G310 manifests ongoing high-mass star formation, as characterised by its associated, centrally-located luminous IRAS source
I13484-6100 with $L_{\rm bol} \sim 10^{4.8}$\,\lsun\ \citep{Bro96}. The presence of a cometary ultra-compact (UC){\Hii} region (G310.1420+00.7583B, \citealt{Pur18}), detection of maser emission from different species such as $\rm OH$, $\rm H_2O$ and $\rm CH_3OH$ ( 6.7\,GHz, Class\,II) masers \citep{Wal98,Urq09,Gre12}, an extended green object (EGO, \citealt{Cyg08}), and radio jets \citep{Pur18} makes the G310 HFS cloud an ideal target for  addressing the mechanisms involved in high-mass star formation by studying the hierarchical kinematics and dynamics covering different scales from clouds to cores. 

\section{Observations} \label{sec:obs}
The hub 
of the G310 HFS cloud (see Fig.\,\ref{fig:overview}) was observed 
as part of the ATOMS (ALMA Three-millimeter Observations of Massive Star-forming regions; \citealt{Liu20a,Liu20b,Liu21}) survey. Combined 7m+12m continuum and line emission data at 3\,mm are used here, which offers a field of view of $\sim 80\arcsec$ or 2.1\,pc at the distance of G310, and a maximum recoverable scale of  $\sim 60 \arcsec$ or 1.6\,pc.
We utilize HCO$^{+}$ and H$^{13}$CO$^{+}$~(1-0) spectral line data to probe the kinematics and dynamics of dense gas and the SiO~(2-1) and CS~(2-1) transitions for tracing the shocked gas.
The synthesized beam size is $2.7\arcsec \times 2.0\arcsec$ for continuum and \htcop spectral line and $3.1\arcsec \times 2.3\arcsec$ for the other spectral lines investigated here. These values correspond to 
0.06\,pc and 0.07\,pc at the distance of G310. In addition, the data sensitivity is $\sim 0.3$\,mJy~beam$^{-1}$ for continuum, and $\sim$~[12, 8, 3]\,mJy~beam$^{-1}$ for the [HCO$^{+}$, H$^{13}$CO/SiO, CS] lines at a native velocity resolution of [0.1, 0.2, 1.5]\,\vel. 

Additionally,
{\it Spitzer} 3.6, 4.5 and 8.0\,\um\ images retrieved from the GLIMPSE survey \citep{Ben03} are used to facilitate the identification of the HFS morphology of G310 (Fig.\,\ref{fig:overview}).
These images have angular resolutions better than 2\arcsec\ \citep{Ben03} that nearly match the resolution of the ATOMS data used here. Further, to explore the dynamics (i.e., velocity gradients and infall motions) of the cloud and clump scales, we use the $3\arcmin \times 3\arcmin$--sized maps of HCO$^{+}$ and H$^{13}$CO$^{+}$ (1-0) obtained from the Millimetre Astronomy Legacy Team 90\,GHz (MALT90) survey \citep{Jac13,Fos11,Fos13}. These data have an angular resolution of 38$\arcsec$ or 1.0\,pc, and a sensitivity of 0.25\,K at a spectral resolution of 0.11 km s$^{-1}$ \citep{Jac13}.   

\section{Results and analysis}
\label{sec:analysis}

\begin{table*}
\centering
\caption{Physical properties of the central clump and the central core in HFS G310}
\label{tab:clump core}
\resizebox{17cm}{!}{
\begin{tabular}{cccccccccc}
\hline\hline
name                 &$\rm R.A.$   &  $\rm Dec.$  &  Radius     &  $\rm T_{dust}$  &  $\rm Mass$        & 	  $\rm \alpha_{vir}$  	&   $\rm {V_{infall}}$	        &      $\rm \dot M_{infall}$                  &      Data source     \\
                     &             &               & $(\rm pc)$  &    $\rm K$	   &  $(\rm M_{\odot})$ &                          &   $(\rm km~s^{-1})$           &    $\rm (10^{-3} \times M_{\odot} yr^{-1})$ &                \\
\hline
Central Clump        & 13:51:57.76 & -61:15:43.72  & 0.6 	     &    31.8          &  $\rm 1280\pm260$ & 	$\rm 0.4\pm0.1$ 	&    $\rm 1.1\pm0.6$           &          $2.4\pm1.7$                       &      Mopra     \\
$\rm Central~Core$ & 13:51:58.30 & -61:15:41.40  & 0.036 	     &    100           &  $\rm 106\pm21$     &  	$\rm 1.5\pm0.6$ 	&    $2.1\pm0.1$             	&          $6.5\pm1.3$                         &     ATOMS      \\
\hline
\end{tabular}
}
\begin{flushleft}
Notes: Col.1: source name; Col.2: radius of clump or core; Col.3: dust temperature (see text in Sect.\,3.1); Col.4: mass; Col.5: viral parameter; Col.6: infall velocity; Col.7: mass accretion rate.
\end{flushleft}
\end{table*}

\begin{figure*}[t]
    \centering
    \includegraphics[angle=0, width=0.95\textwidth]{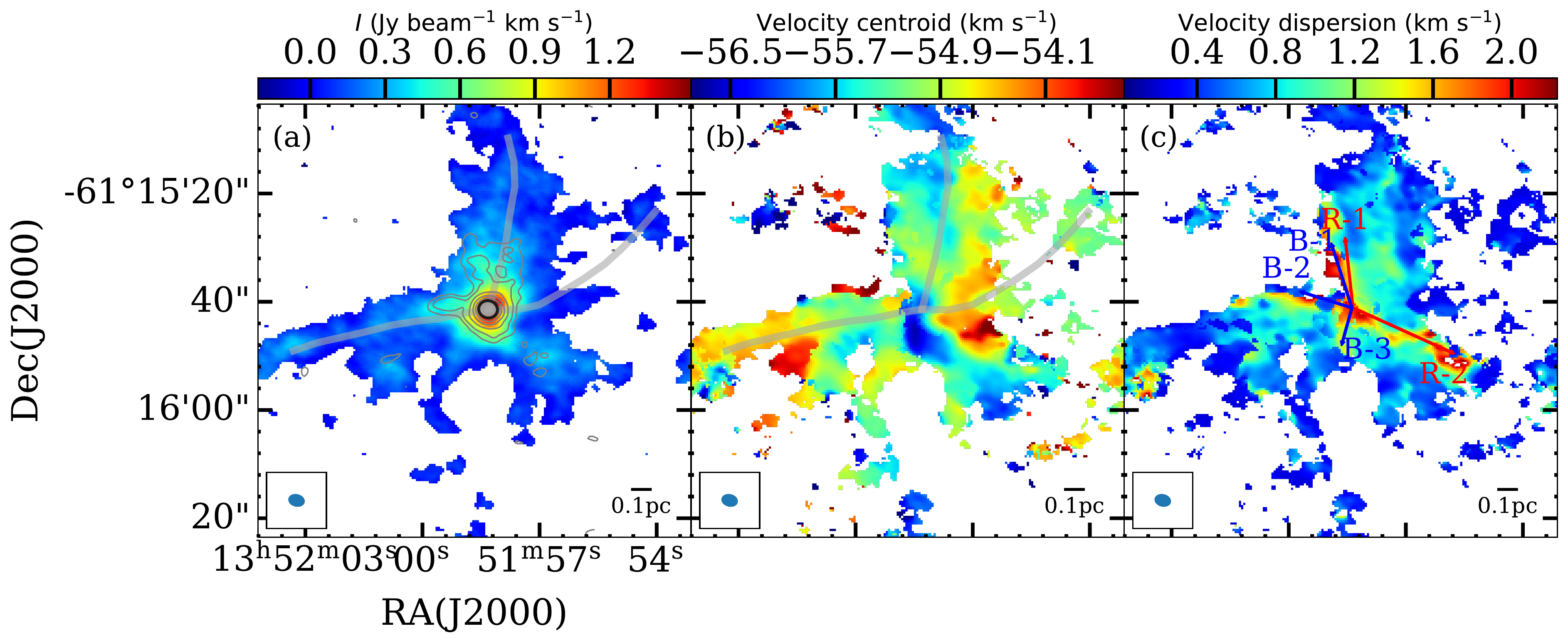} 
    \caption{Moment maps of \htcop~(1--0) from the ATOMS data for the central hub region of G310. (a:) integrated-intensity (Moment\,0) map overlaid with the ATOMS 3\,mm continuum contours. The contour levels follow the same trend as that in Fig.\,\ref{fig:overview}. The central massive core is indicated in grey ellipse with a black edge. (b:) mean velocity (Moment\,1) map. The grey thick line show the filaments of G310 HFS cloud in panels a, b.
    (c:) velocity dispersion (Moment\,2) map. The blue and red arrows show the blue- and red-shifted outflow lobes (i.e., R-1/2 and B-1/2/3), respectively. The ALMA beam size is shown in the bottom left.} 
    \label{fig:h13co}    
\end{figure*}

\subsection{Continuum emission}\label{subsec:result:cont}
The HFS morphology seen in the mid-infrared is also revealed in the 
sub-mm and mm dust emission maps. The
870\,\um\ dust emission obtained from the ATLASGAL survey \citep{Sch09} is shown in Fig.\,\ref{fig:overview}, where the signature of filaments F1 and F2 extending outwards the centrally-located, massive clump is clearly discernible. The higher resolution ATOMS 3\,mm dust emission contours also trace the inner part of these filaments which originate from the central massive core (see Fig.\,\ref{fig:h13co}a). 

We estimate the mass of the ATLASGAL clump following Eq.\,B1 of \citet{Liu21}. We define a circular region of radius 0.6\,pc (white circle in Fig.\,\ref{fig:overview}) which encloses most of the 870\,\um\ emission ($\sim 8.7$\,Jy). 
Assuming a dust opacity of 0.0178\,cm$^2$~g$^{-1}$ that accounts for a gas-to-dust gas mass ratio of 100 \citep{Oss94}, clump mass is estimated to be 
1,280$\pm260$\,\msun\ for a given clump-average temperature of 31.8\,K \egcite{Bro96, Fau04}, where the error comes mostly from the uncertainty in the distance estimate. The mass of the candidate hot molecular core (HMC, \citealt{Liu21}), i.e., central massive core (radius of 0.036~pc), mass is estimated to be 106$\pm21$\,\msun\ at an assumed temperature of 100\,K. Given the non-detection of H40$\alpha$ line emission from the central massive core \citep{Liu21}, the 3\,mm continuum can be considered to be mostly from thermal dust emission with negligible contamination from ionized gas (free-free emission). The estimated parameters for the central massive clump and the central core are listed in Cols.\,4--6 of Table\,\ref{tab:clump core}.

\begin{figure*}
   \centering
   \includegraphics[angle=0, width=0.42\textwidth]{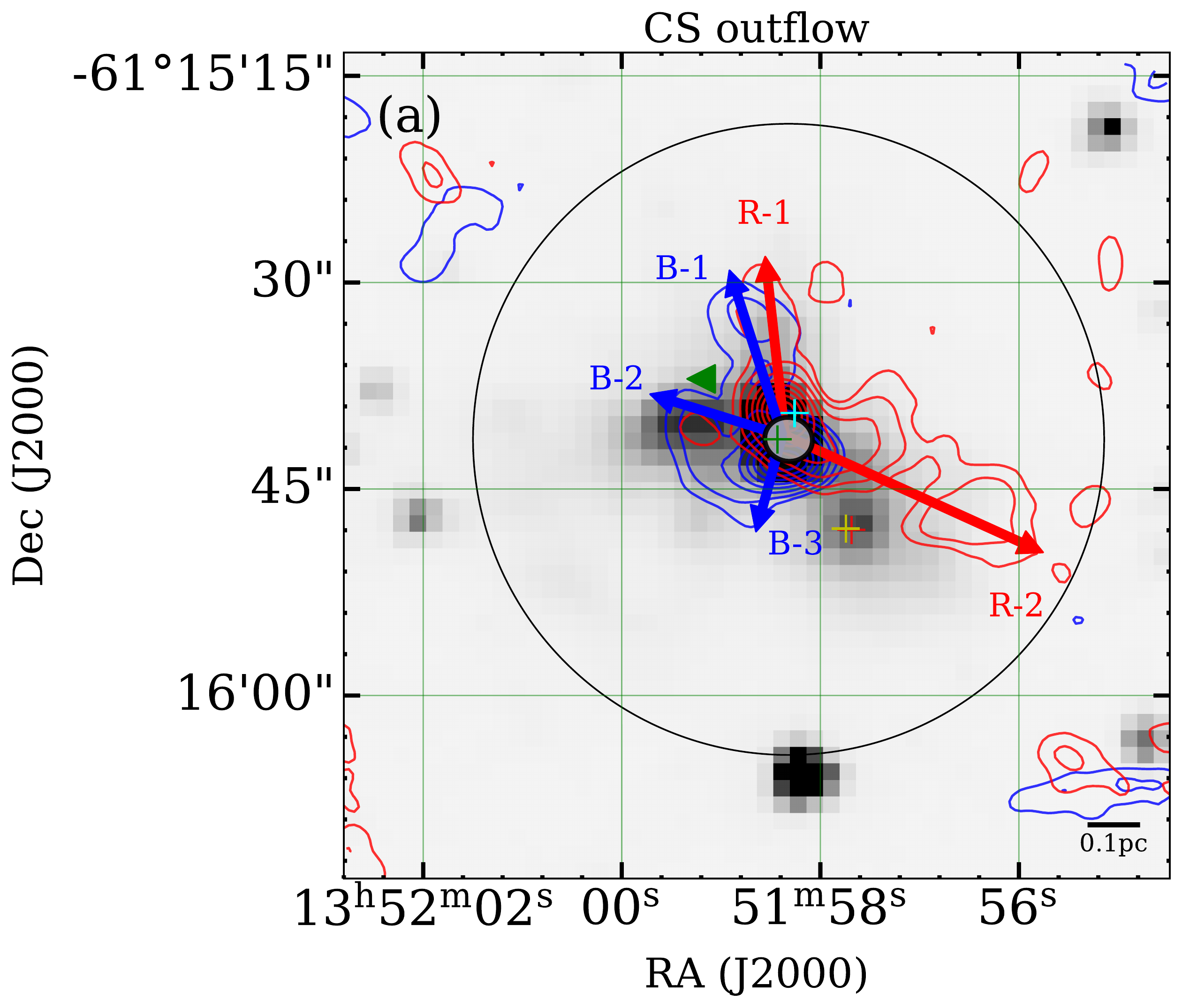} 
   \includegraphics[angle=0, width=0.4\textwidth]{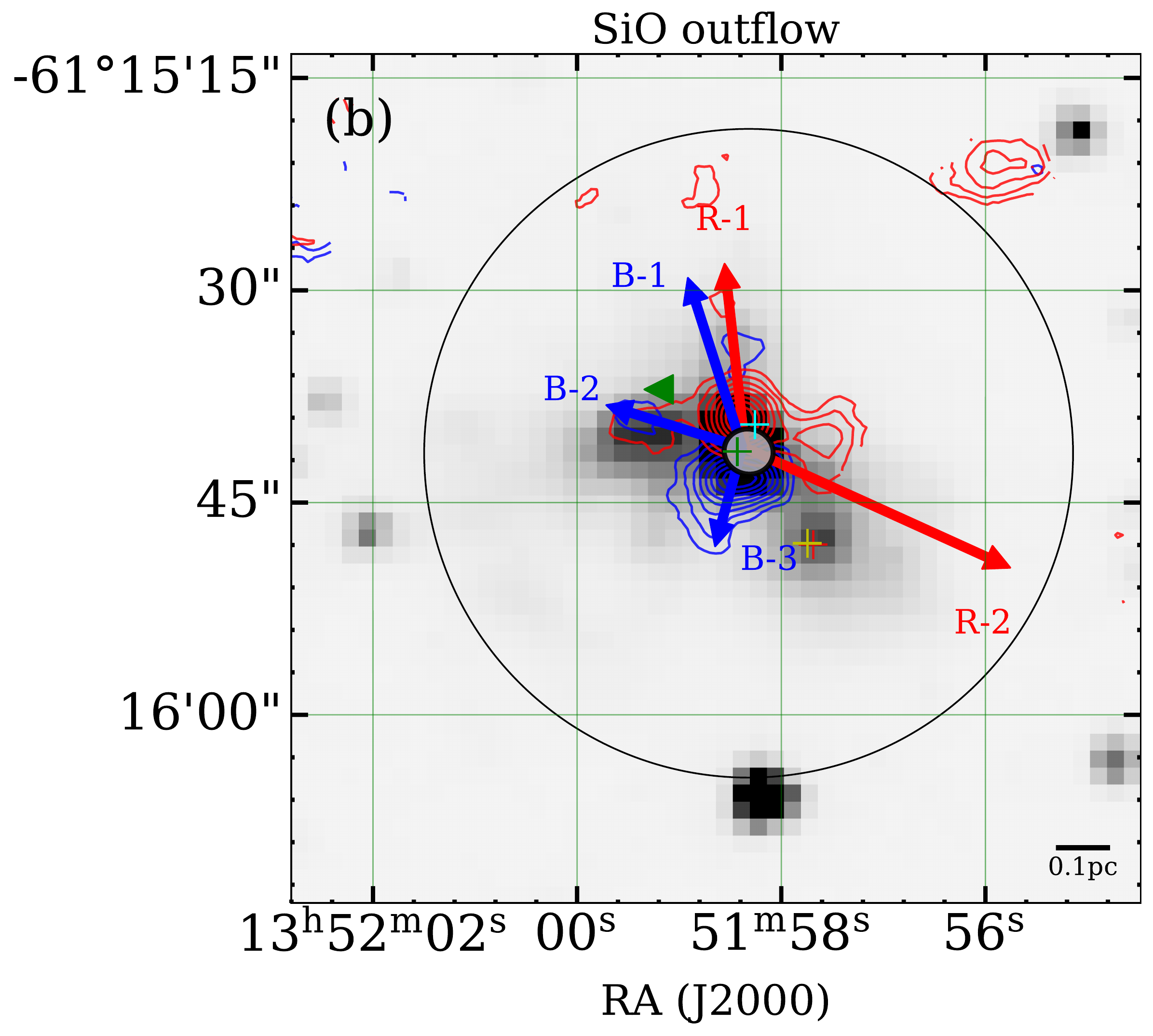} 
    \caption{Panel (a): CS (2--1) outflows overliad on the {\it Spitzer} 4.5\,\um\ map. The contour levels start at 3 rms ($\sim 0.15$\,Jy\,beam$^{-1}$\,\vel), increased by steps following the same power law as in Fig.\,\ref{fig:overview}. Panel (b): SiO (2--1) outflows overlaid on the {\it Spitzer} 4.5\,\um\ map, the contour same with Panel (a), but start at 3 rms ($\sim 0.08$\,Jy\,beam$^{-1}$\,\vel). In both panels, the blue and red arrows show the blue- and red-shifted outflowing lobes (i.e., R-1/2 and B-1/2/3), respectively; The 3\,mm central massive core is shown as a grey ellipse; The black circle represents the central massive clump of radius 0.6\,pc; The green triangle gives the location of the identified EGO. The red cross shows the location of the \uchii\ region, the green, cyan and yellow crosses represent the $\rm CH_3OH$, OH and $\rm H_2O$ maser, respectively.
    }
   \label{fig:outflows}
\end{figure*}

\subsection{\htcop~(1--0) line emission} \label{subsec:result:molecular}

\htcop\ line emission is generally considered  optically thin \citep{Liu22a,Liu22b} and hence, a good tracer of kinematics/dynamics and density structures of different scales. In the analysis presented, the field of view of ATOMS is considered as the central hub of the G310 HFS. The average spectrum of \htcop~(1--0) over the central hub region is single-peaked from which the systemic velocity is estimated to be $V_{\rm lsr}=-55.2$\,\vel.
\htcop~(1--0) moment maps, generated from a de-noised/modelled \htcop\ data cube (see Appendix\,\ref{app:denoise_moms}), are shown in Fig.\,\ref{fig:h13co}. 
As seen in the Moment\,0 map, the spatial distribution of \htcop~(1--0) is much more extended compared to the 3\,mm dust emission which is confined to the innermost region. Two of the overlaid (in grey) filament skeletons (F1 and F2) traced from the 8.0\,\um\ image are discernible in the \htcop~(1--0) emission (above 5\,rms). 
An elongated gas emission is seen from the central core in the south-west direction. This is also visible as an IR-bright filament-like structure in the 8.0\,\um\ emission (see Fig.\,\ref{fig:overview}). As seen in Fig.\,\ref{fig:h13co}(c) and discussed later, this feature is co-spatial with the outflow R-2 and hence likely associated with shocks from the outflow \citep{DeB17}.

The velocity field, displayed in the Moment\,1 map, shows gradients along the filaments (i.e., F1--F3). 
In a later section (see Sect.\,\ref{subsec:result:vg}), a detailed analysis of the filament velocity structure will be presented. Interestingly, a very local, radially symmetric velocity gradient can be found across the IR-bright filament discussed above. 

The observed pattern of the velocity gradient can originate from various physical processes like, 

disk rotation, collision of local gas flows, and/or stellar feedback such as outflows and winds. The possibility of disk rotation can be ruled out since no Keplerian-like signature is found in the position-velocity (PV) diagram particularly made along the velocity
gradient orientation (not presented in the paper). Although the possibility of the collision between gas flows cannot be completely rejected, a more plausible explanation is the outflow feedback since the red/blue-shifted outflow lobes (see Fig.\,\ref{fig:outflows}, and Sect.\,\ref{subsec:result:outflows}) 
are found to be spatially coincident with the red/blue-shifted velocity components of the local velocity gradient.

The Moment\,2 map (see Fig.\,\ref{fig:h13co}c) shows the velocity dispersion to be non-uniform with higher values in the inner region, 
implying active star-forming activity (e.g., outflows). Regions with high velocity dispersion appear as a slingshot-like pattern co-spatial with the identified outflow lobes (see Fig.\,\ref{fig:h13co}c).
Quantitatively, the velocity dispersion of the entire hub region has a median value of 0.58\,\vel\ ranging from 0.22--2.20\,\vel. Following \cite{Liu19},
the typical value of the total turbulent velocity (including thermal and non-thermal contribution) is estimated to be 0.66\,\vel\ taking the thermal speed as 0.34\,\vel\ for a given clump-averaged temperature of 31.8\,K. This result suggests that the central hub region of the G310 HFS cloud is overall supersonic.

\begin{figure*}
    \centering
    \includegraphics[angle=0, width=0.85\textwidth]{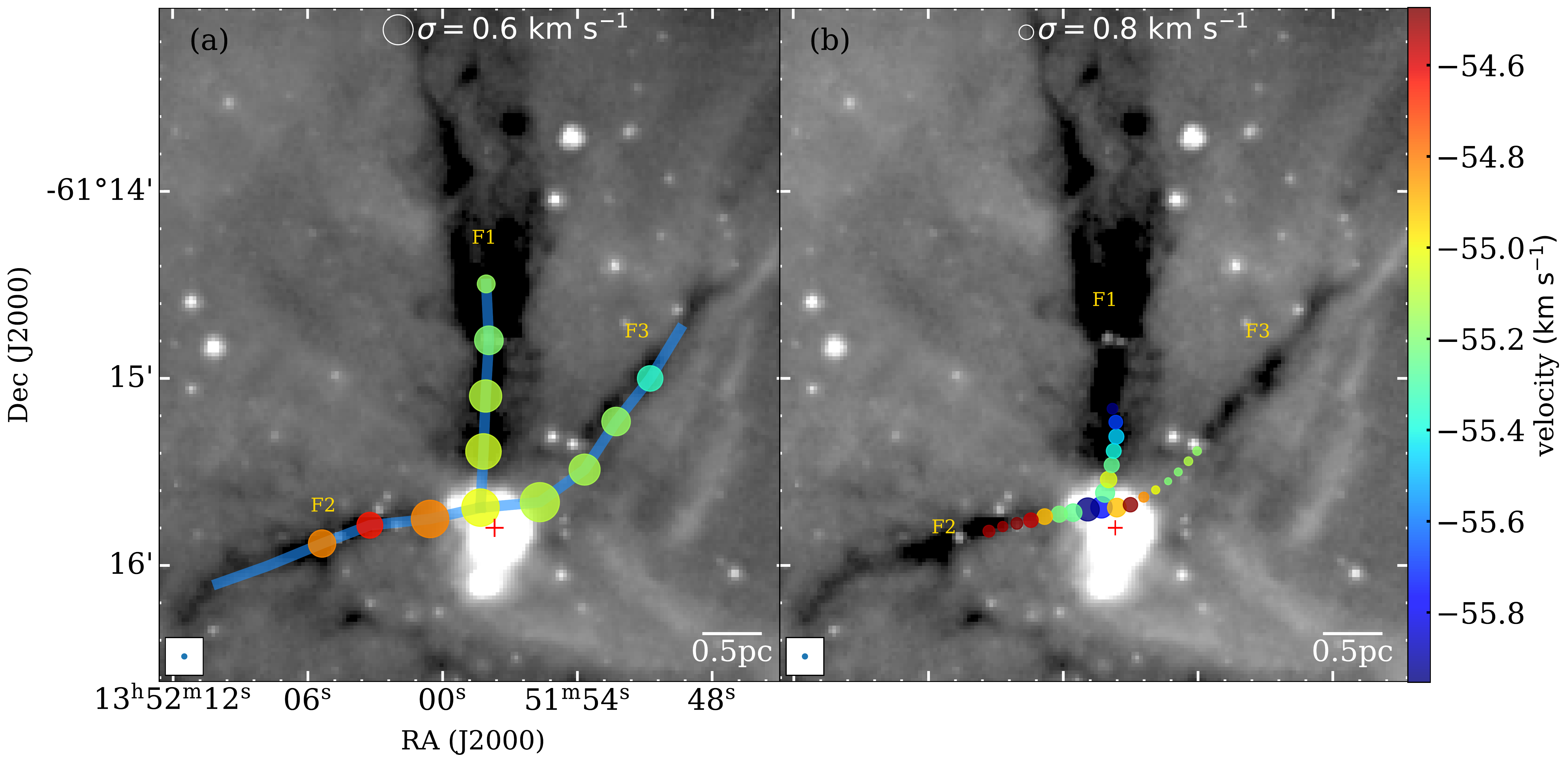} 
    \caption{ Distribution of the velocity and its dispersion along filaments derived from \htcop~(1--0) of the MALT90 data (panel\,a) and from the same line of the ATOMS data (panel\,b). In both panels, the grayscale is the {\it Spizter} 8{\um} map. The color and size of the filled circles represent the velocity, and its dispersion, respectively. 
    The blue thick lines show the filaments of the G310 HFS cloud (in panel\,a).
    The white empty circle in the upper middle of each panel indicates the median velocity dispersion of all measurements (circles), i.e., 0.6 $\rm km~s^{-1}$ in panel\,a and 0.8 $\rm km~s^{-1}$ in panel\,b. The beam size of the 8\,\um\ map and the 0.5\,pc scalebar are shown at the bottom left and right corners, respectively. }
   \label{fig:line width}
\end{figure*}

\subsection{Molecular outflows}\label{subsec:result:outflows}
Two shocked gas tracers, CS~(2--1) and SiO~(2--1), are used for reliable identification of outflow lobes in the G310 HFS. From the line wings of the average spectrum of each transition over the central hub with significant emission (above 5\,rms), the blue/red-shifted outflowing gas is integrated over [-75, -60]/[-50, -35]\,\vel\ for CS~(2--1), and [-85, -60]/[-50, -25]\,\vel\ for SiO~(2--1).

\begin{table*}
    \centering
    \caption{Physical properties of hub-composing filaments in G310}
    \label{tab:filament}
    \resizebox{17cm}{!}{
    \begin{tabular}{cccccccc}
    \hline\hline
    name          &     Area 		        &     Length   &  Width  &   	 $\rm M_{fil}^a$      & 	  $\rm M_{line}$  		            &   $\rm{\nabla_{fil}}^b$	        &      $\dot M_{||}$                           \\
                  &     $(\rm pc^2)$        &    $(\rm pc)$ & $(\rm pc)$ &     $(\rm M_{\odot})$  &     $(\rm M_{\odot} pc^{-1})$ 	  & $(\rm km~s^{-1}~pc^{-1})$         &    $\rm (10^{-4} \times M_{\odot} yr^{-1})$   \\
    \hline
    Filament1	  &  	 7.3 	&     	5.8     &  1.3  &   	 $3566\pm713$       & 	$620\pm124$ 	                    &    $0.10\pm0.02$             	&          $3.6\pm1.5$       \\
    Filament2	  &	    2.4 	&	    3.3     &  0.7  &	     $1353\pm271$       &	$409\pm82$                          &	 $0.17\pm0.14$         	    &          $2.4\pm2.4$        \\
    Filament3	  &	    3.5 	&	    4.6     &  0.8  &	     $777\pm155$        &	$167\pm34$                          &	 $0.18\pm0.04$  		    &          $1.4\pm0.6$        \\
    \hline
\end{tabular}
}
\begin{flushleft}
Notes: Col.1: name; Col.2: area; Col.3: length; Col.4: width; Col.5: mass; Col.6: mass-per-unit-length of filament; Col.7: filament-aligned velocity gradient which is measured from the MALT90 data (see Fig.\,\ref{fig:vel:grad}a); Col.8: mass inflow rate.
\end{flushleft}
\end{table*}

Five outflow lobes are identified and 
presented in Fig.\,\ref{fig:outflows}a for the CS tracer and Fig.\,\ref{fig:outflows}b for SiO. 
These lobes show similar structure in both tracers though the spatial extent in CS emission is comparatively larger. 
 
It is possible that faint, extended SiO emission is below the sensitivity of the observations (about 0.2\,K). 

The identified outflows in both tracers are consistent with detection of radio jets reported in Fig.\,4 of \cite{Pur18}, and the location of the associated EGO (\citealt{Cyg08}, see Fig.\,\ref{fig:outflows}a).
Of particular interest is the spatial extension of the outflowing lobe, R2, oriented along the IR-bright filament discussed earlier.

Moreover, all outflowing lobes appear to originate from the most massive central core, suggesting that the core is forming high-mass stars in cluster.

\begin{figure*}
    \centering
    \includegraphics[angle=0, width=0.45\textwidth]{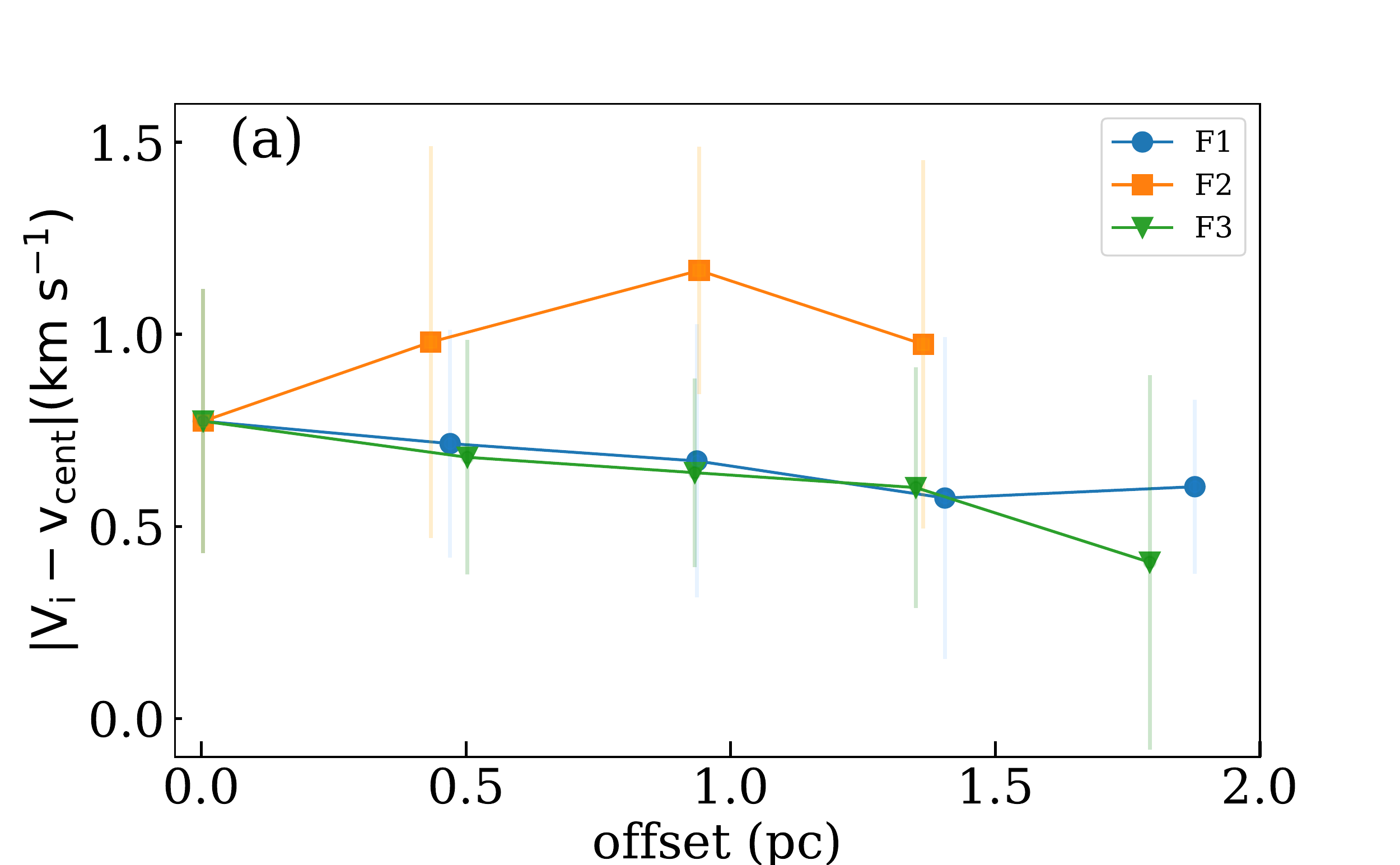} 
    \includegraphics[angle=0, width=0.45\textwidth]{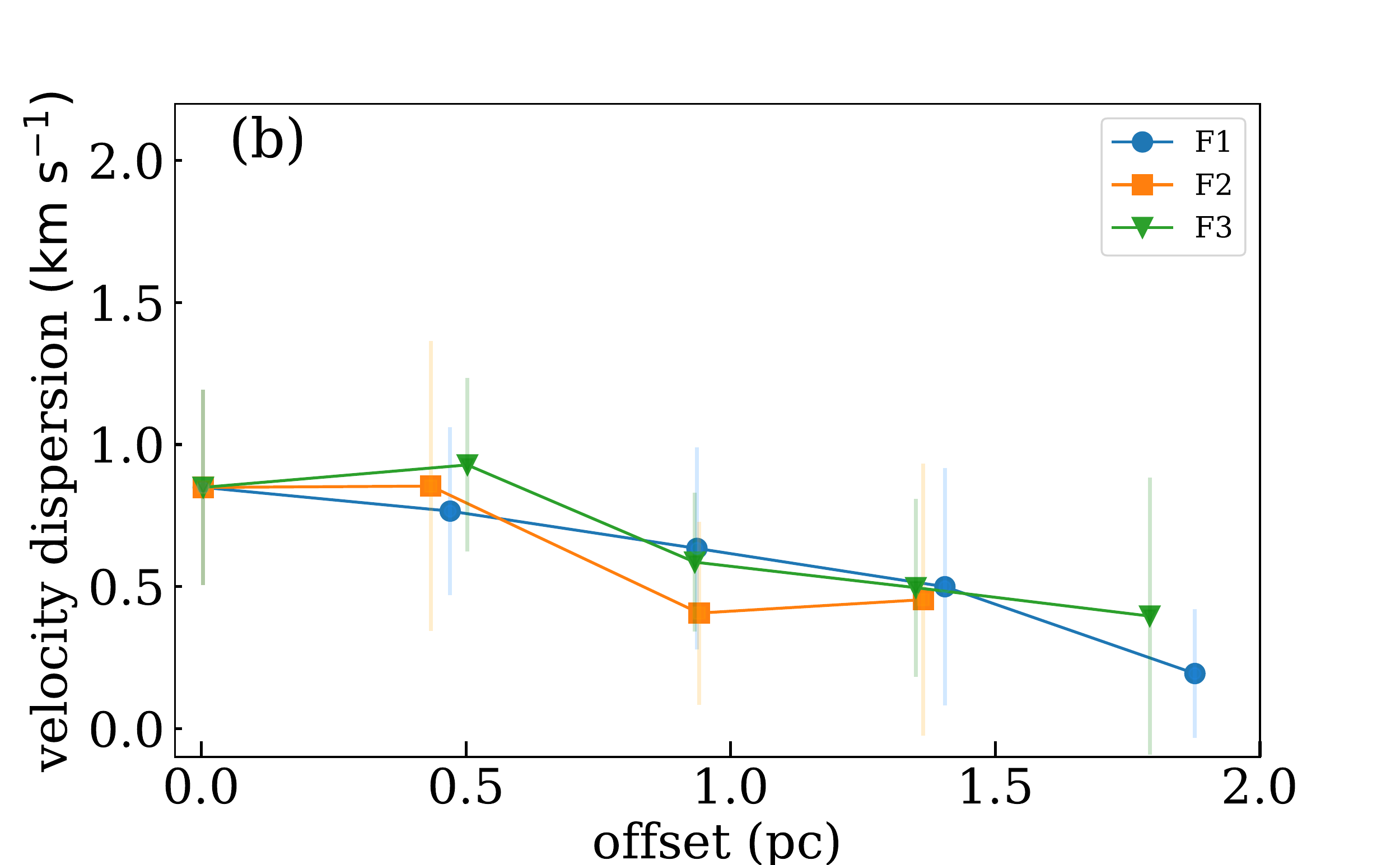} 
    \includegraphics[angle=0, width=0.45\textwidth]{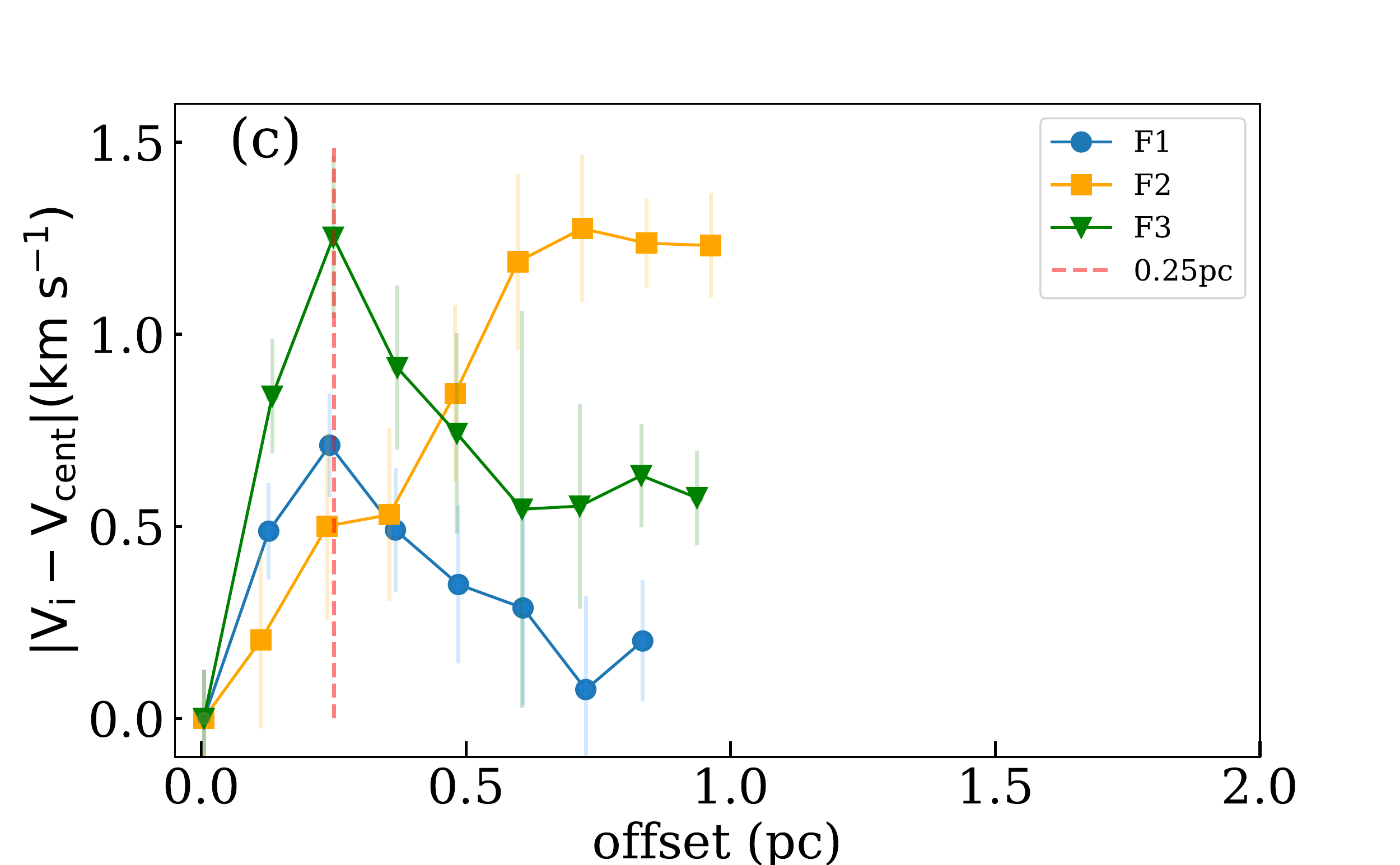} 
    \includegraphics[angle=0, width=0.45\textwidth]{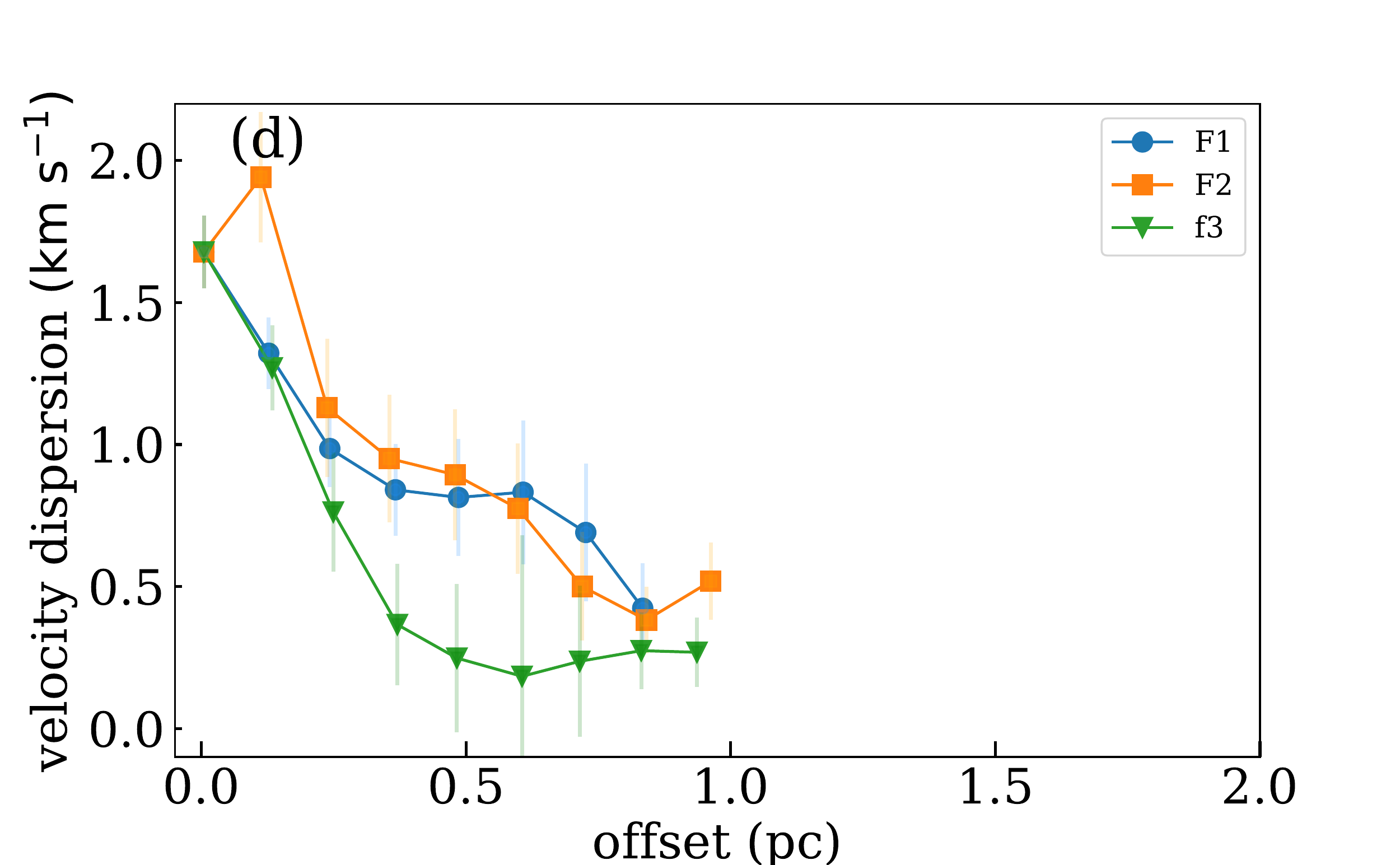} 
    \caption{ (a) Relative velocity of \htcop~(1--0) from the MALT90 data for three filaments as a function of offset from the gravitational potential well centre (the central massive core). (b) same as panel\,a but for the velocity dispersion. (c) same as panel\,a but for the \htcop~(1--0) line from the ATOMS data. The vertical dashed line indicates the offset at which 
    the velocity gradient trends break. (d) same as panel\,c but for the velocity dispersion. Note that in panels\,(a) and (c), the same $V_{\rm centre}=-55.2$\,\vel\ is used for comparison.}
   \label{fig:vel:grad}
\end{figure*}

\subsection{Multi-scale velocity gradients} \label{subsec:result:vg}
The cloud and clump scale velocity fields are probed using the \htcop~(1--0) line emission data from the MALT90 and ATOMS surveys, respectively. The filament dynamics is retrieved by fitting single Gaussian profiles to the observed spectra retrieved over circular apertures. For the MALT90 data, these apertures are over half beam and for the ATOMS data they are over one beam. 

Investigating the cloud scale velocity structure, Fig.\,\ref{fig:line width}a presents the distribution of the derived velocity parameters from the MALT90 data overlaid on {\it Spitzer} 8\,\um\ emission, where the color and size of the circle encodes the velocity, and its dispersion, respectively. Longitudinal velocity gradients are clearly evident along the three hub-composing filaments and the velocity dispersion is seen to increase toward the centre. The red- and blue-shifted velocities, with respect to the systemic velocity of the HFS cloud, indicate filament-rooted gas motions toward us for F1 and F3 and away from us for F2. This is further emphasized in Fig.\,\ref{fig:vel:grad}a. A monotonic trend is seen for the filaments within measured velocity uncertainties. The trend is decreasing toward the centre, for F2 but increasing for both F1 and F3. 
Note that the coherent trend seen in F1 suggests that the subfilaments (if any) in F1 have the nearly same velocity gradient, and thus can be treated as a single entity for kinematic analysis.
The velocity gradients along the three filaments are measured from Fig.\,\ref{fig:vel:grad}a to be $0.1\pm0.1$\,\vel~pc$^{-1}$ for F1, $\sim 0.2\pm0.1$\,\vel~pc$^{-1}$ for F2 and $\sim 0.2\pm0.1$\,\vel~pc$^{-1}$ for F3, which are presented in Table\,\ref{tab:filament}.

The clump-scale velocity gradient is discerned from the ATOMS data and shown in Fig.\,\ref{fig:line width}b. The direction of gas motion in the filaments and the velocity dispersion trend seen at the cloud scale are retained at the scale of the clump suggesting the same origin for both. However, the gradients are seen to be steeper than at the cloud scale. For filaments F1 and F3, a reversal is seen around 0.25\,pc with a sharp increase in the velocity dispersion as well. This is likely the result of strong stellar feedback as the gas kinematics and dynamics of the inner region of the clump will be strongly influenced by the central, luminous IRAS source. The amplitudes of the velocity gradients below and above 0.25\,pc in sequence are estimated to be $3.0\pm0.6$ and $\rm 0.9\pm0.2$\,\vel,pc$^{-1}$ for F1, $2.2\pm0.1$ and $\rm 1.2\pm0.3$\,\vel\,pc$^{-1}$ for F2, and $5.2\pm0.8$ and $\rm 0.9\pm0.3$\,\vel\,pc$^{-1}$ for F3. 
It is worth noting that 
 the dynamical parameters derived from the MALT90 and ATOMS data are different (Fig.\,\ref{fig:vel:grad}), where 
 the velocities derived from the former are around 3--4\,\vel\ blue-shifted than those from the latter while the maximum velocity dispersion derived from the former is around 2 times higher than those from the latter. This could be attributed to the different beam sizes of the two data sets, where the relatively poor beam (38\arcsec) of the MALT90 data samples more cold quiescent envelope gas (mostly likely blue-shifted) compared to that probed with
 the finer beam ($\sim2$\arcsec) of the ATOMS data. These differences result in much shallower velocity trends on cloud scale than on clump scale (i.e., Fig.\,\ref{fig:vel:grad}a v.s. Fig.\,\ref{fig:vel:grad}c), but
 do not affect the general velocity trends as analysed above.

\subsection{Multi-scale gas infall/accretion } 
\label{subsec:result:inflow}
The observed cloud-scale velocity gradients in G310 can be considered as imprints of the filament-rooted inertial inflow onto the gravitational potential well of the HFS cloud (see below). Following \citet{Kir13}, the parallel mass inflow rate ($\dot{M_{\rm fil}}$) can be estimated by assuming a cylindrical filament of mass $M_{\rm fil}$, and length $L_{\rm fil}$ at an inclination angle of $\phi$ with respect to the plane of the sky ($45\degr$ used here). The total $\dot{M_{\rm fil}}$ of the hub-composing filaments is estimated to be $0.7\pm0.5\times10^{-3}$\,\mdotyr\ (see Table\,\ref{tab:filament}, and Appendix\,\ref{app:accretion_rates} on the detailed calculation for each individual filaments). This value should be a lower limit, and thus can be above $\sim10^{-3}$\,\mdotyr\ if the remaining gas inflowing off the filaments within the cloud is considered \citep{Per13}.

\begin{figure}
   \centering
   \includegraphics[angle=0, width=0.5\textwidth]{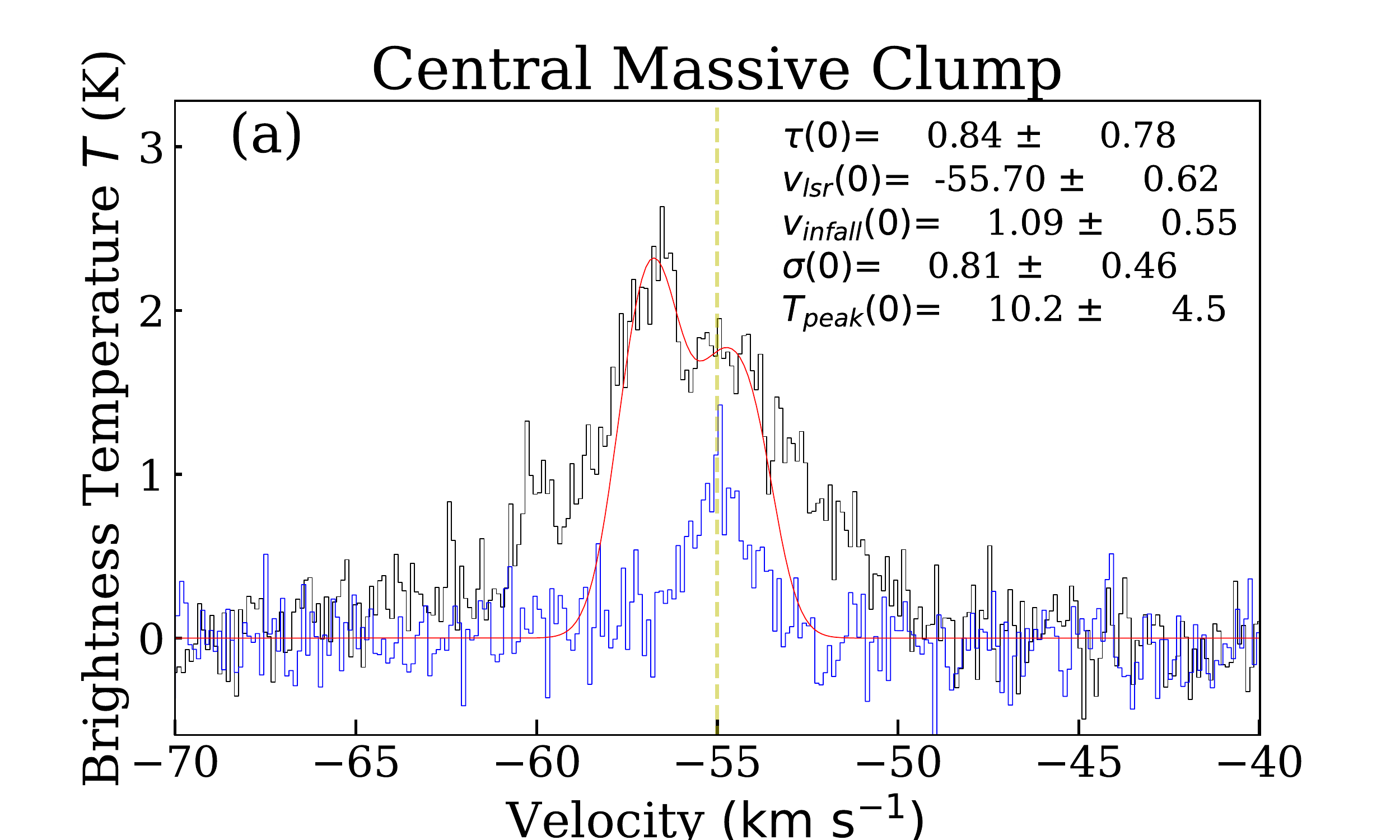} 
   \includegraphics[angle=0, width=0.5\textwidth]{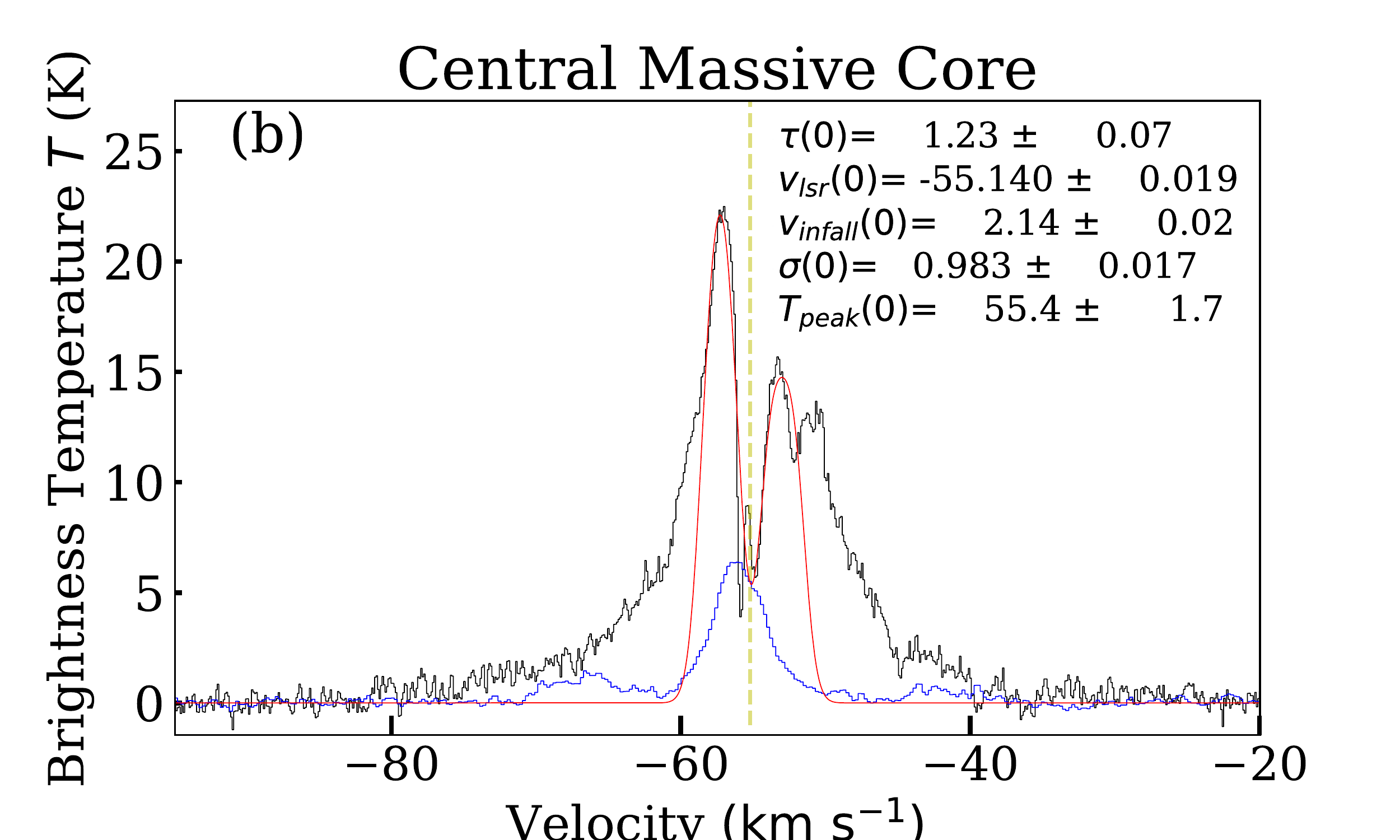} 
    \caption{ Average spectra of J=1--0 of \hcop\ (in black) and \htcop\ (in blue) for the massive clump (panel\,a), and central massive core (panel\,b).  The red curve represents the Hill5 model fitting while the yellow vertical line indicates the systematic velocity derived from the optically-thin \htcop\ line.  The spectra for the clump and core are extracted from the MALT90, and ATOMS data, respectively. The model-fitted parameters are given  at the upper right corner of each panel.}
   \label{fig:infall}
\end{figure}

The classical infall signature of a blue asymmetric profile of the optically thick \hcop~(1--0) transition \citep{Liu16} is observed for the G310 clump from the MALT90 data and displayed in Fig.\ref{fig:infall}. This infall signature allows to attribute the same trend of the cloud and clump scale velocity gradients to the same mass infall/inflow origin.
The infall velocity is estimated using the radiative transfer model ``Hill5" \citep{De05}. The fitted model is shown in Fig.\ref{fig:infall}a. The estimated value of $1.1\pm0.6$\,\vel, is in good agreement with those reported for clumps \egcite{Tra18}. Subsequently, the clump-scale mass infall/accretion rate \egcite{San10},
$\dot{M} = M_{\rm clp} V_{\rm clp}^{\rm in} / R_{\rm clp}$ is estimated to be $(2.4\pm 1.7)\times10^{-3}$\,\mdotyr\ , where $M_{\rm clp}$ is the enclosed mass within $R_{\rm clp}=0.6$\,pc and $V_{\rm clp}^{\rm in}$ is the infall velocity at the radius $R_{\rm clp}$ (see Table\,\ref{tab:clump core}). At the core level, the ATOMS \hcop~(1--0) transition shows similar infall signature (see Fig.\ref{fig:infall}b). This finds support in the detected outflows associated with the central massive core which is evidence of core-scale gas infall on to the embedded protostar(s). Using the same approach discussed above, the infall velocity, $V_{\rm core}^{\rm in}$ and the core mass accretion rate is estimated to be $(2.1\pm0.1)$\,\vel\ and $(6.5\pm 1.3)\times10^{-3}$\,\mdotyr, respectively, the later being typical of high-mass star-forming cores \citep{Lop10,Sah22}. The above derived parameters for both central clump and core are given in Cols.\,8--9 of Table\,\ref{tab:clump core}.

The clump- and core-scale infall are consistent with their subvirial state. The virial parameter, $\alpha_{\rm vir}=\frac{5\sigma^2_{\rm turb} R}{GM}$ \egcite{Kru05,Liu19} where G is a gravitational constant. The turbulent velocity  $\rm \sigma_{turb}\sim (0.9\pm0.4)$\,\vel\ is derived from the clump-averaged \htcop~(1--0) spectrum of MALT90 survey. For the turbulent central massive core, $\sigma_{turb}\sim (2.0\pm0.1)$\,\vel\ derived from \htcop~(1--0) of the ATOMS data. Using these values along with the estimated radius and mass (see Sect.\,\ref{subsec:result:cont}), the virial parameter, $\alpha_{\rm vir}$, is calculated to be $\sim 0.4\pm0.1$ and $\sim 1.5\pm0.6$ for the clump and the core (listed in Col.\,7 of Table\,\ref{tab:clump core}), respectively, satisfying at both scales $\alpha_{\rm vir}^{\rm crit}\leq2$ for gravitational collapse \egcite{Kau13}.

\section{Discussion}
\label{sec:discussion}
\subsection{Multi-scale scenario of high-mass star formation}
Many observational and numerical studies are gradually converging to a multi-scale mass accretion scenario for the formation of high-mass stars, and these highlight the role of HFSs in high-mass star formation \citep{Beu18,V19,Pad20,Kum20,Liu22b,Sah22,Liu23}. HFSs have been the focus of several statistical studies \egcite{Kum20, Zho22, Liu23}. In particular, \cite{Liu23} studied a sample of 17 HFSs composed of two distinct evolutionary stages using high-angular resolution ($\sim1$--2\arcsec) ALMA 1.3\,mm \citep{Li23,Mor23} and 3\,mm continuum data \citep{Liu20b,San19}. \citet{Liu23} discuss the observed trend on multi-scales (i.e., clumps and cores) of increasing mass and mass surface density with evolution from IR-dark to IR-bright stage, the mass-segregated cluster of YSOs, and the potentially preferential escape directions of outflow feedback. These observed facts strongly advocate for multi-scale mass accretion/transfer as the major agent for high-mass star formation in the HFSs.

For G310 HFS, the smooth large-scale (cloud-scale), filament-aligned coherent velocity gradients could be due to the effect of either gravity-driven gas motion towards the central massive core, as described in the GHC model, or turbulence-driven large-scale inertial inflow discussed in the I2 model. Both models claim to facilitate mass accumulation in the hub and thus high-mass star formation therein. In the central hub region of the G310 cloud covered by the ATOMS data, the gas infall is evident through the filament-rooted velocity gradients present all the way down to the HFS centre where high-mass stars form. At the core-scale, the ongoing gravitational collapse results in mass accretion onto the centrally-embedded luminous protostars.

The observed morphology of G310 HFS suggests the hierarchical density structures (i.e., clouds, clumps and cores) to be spatially connected and dynamically linked through a multi-scale, dynamical process of mass accretion. Observed in other high-mass star forming HFS clouds \egcite{Liu12,Per13,Kum20,Per14,Avi21,San21,Liu22a,Liu22b}, this dynamical mass accretion process has been highlighted in the theoretical models such as GHC and I2. 
Of particular mention are the two most-studied high-mass HFS clouds, SDC335 \citep{Avi21,Olg21,Olg22,Xu23} and G34 \citep{Liu22a}, which give detailed observational evidence in strong support of the theoretical predictions.

For SDC335 HFS, the mass infall rate is similar (i.e., about $10^{-3}$\,\mdotyr) at cloud, clump, and core scales \citep{Avi21,Xu23}. For the G34 HFS, the mass infall rate is around four times higher at clump scale than at core scale. However, the core-scale mass infall rate was indirectly derived from the outflow rate, without accounting for the ionized gas contribution and the exact outflow inclination angle. 
Hence, it is possible
that the core-scale mass infall rate could be underestimated. These studies of SDC335 and G34 suggest that there is a nearly-continuous, multi-scale gas inflow in the early stage of high-mass star formation in HFS clouds. 
From the analysis presented earlier, a similar picture can be conjectured for the G310 HFS cloud.

\subsection{Comparison with theoretical models}
Both GHC and I2 models propose that high-mass stars form through multi-scale, dynamical mass transfer. However, the models differ in how they explain the origin of intermediate-size structures like hub-filament systems. In I2, these structures result from supersonic turbulent flows that are not driven by self-gravity (i.e., “inertial”). In GHC, they originate from large-scale gravitational contraction of the cloud. This implies that I2 requires filaments to be formed by shocks, while GHC predicts a smooth formation process (similar to Bondi flow). In GHC, strong shocks should only occur at the hubs, where the filaments collide. In contrast, I2 predicts that shock tracers should be emitted along the entire length of the filaments. From Fig.\,\ref{fig:outflows}, there is no SiO detection, which traces shocks, along the whole length of the filaments. However, we cannot rule out that expected SiO emission is below the low sensitivity (i.e., 0.2\,K) of our observations. Therefore, the lack of filament-aligned SiO emission does not allow us to favor GHC over I2.

Another key difference between GHC and I2 is that GHC predicts that unbound low-mass or low-column density structures are being compressed by the infall from larger-scale bound structures \citep{Gomez21, Camacho22}. This implies that the virial parameter should decrease with increasing scale in hierarchically embedded structures. In Sect.\,\ref{subsec:result:inflow}, we show that the virial parameter in the G310 HFS decreases by more than a factor of 3 from the core to the clump 
scale. 
However, this could be due to either the compression of the filaments that form the hub, as GHC suggests, or the strong stellar feedback within the central hub’s high-mass star-forming core. To rule out the latter possibility, we need to focus on HFS clouds that are quiet in star formation (e.g., IR-quiet) and perform the same multi-scale analysis of the virial parameter. 
It is, therefore, difficult to favour one model over the other based on the results from this study.

\section{Summary and conclusions} \label{sec:summary}
We present a comprehensive analysis of the hierarchical kinematics and dynamical gas motions associated with a typical hub-filament-system (HFS) cloud, G310, using primarily the 3\,mm continuum observation and molecular line transitions from the ATOMS survey. The cloud- and clump-scale analysis is done with MALT90 data.

Three hub-composing filaments (F1--F3, $5.7\pm1.1\times 10^3$\,\msun\ in total) are detected converging toward the central hub region. The central hub contains a massive clump ($1280\pm160$\,\msun) which harbours a massive HMC, central massive core, of mass 106$\pm21$\,\msun.  Five outflow lobes are identified from the CS~(2--1) and SiO~(2--1) lines in the ATOMS data associated with central massive core, suggesting high-mass star formation in clustered environment. Multi-scale velocity gradients have been revealed from analysis of \htcop~(1--0) of both the MALT90 and ATOMS data. Coherent longitudinal velocity gradients along the filaments coupled with the signature of gravitational collapse in central massive core, suggest continuous, multi-scale gas inflow feeding the high-mass star-forming seed(s) in central massive core. Presence of the luminous ($L_{\rm bol}\sim 10^{4.8}$\,\lsun) IRAS source suggests strong stellar feedback which is supported by the observed reversal of the velocity gradient in the inner (0.25~pc) region of the clump and the associated enhanced velocity dispersion. However, this does not inhibit mass accretion onto the forming central, embedded protostar(s) as evidenced by the subvirial state and infall signature of central massive core, and its associated outflows. 

Our study of a robust 
HFS cloud shows the importance of dynamical, multi-scale mass accretion for the high-mass protostars in the hub of HFS clouds. This supports the multi-scale scenario of high-mass star formation in both gravity-driven (GHC) and turbulence-driven (I2) models. 
Multi-scale gas dynamics need to probed with higher sensitivity observations in a larger sample of HFSs, especially the ones without significant stellar feedback (e.g. IR-dark) to investigate the predictions of these models in detail.

\section{Acknowledgements}

This work has been supported by the National Key R\&D Program of China (No.\,2022YFA1603101).
H.-L. Liu is supported by National Natural Science Foundation of China (NSFC) through the grant No.12103045, and by Yunnan Fundamental Research Project (grant No.\,202301AT070118).
 PS was partially supported by a Grant-in-Aid for Scientific Research (KAKENHI Number JP22H01271 and JP23H01221) of JSPS.
S.-L. Qin is supported by NSFC under No.12033005.
T. Liu acknowledges the supports by NSFC through grants No.12073061 and No.12122307.
Ke Wang acknowledges support from the National Science Foundation of China (11973013), the China Manned Space Project (CMS-CSST-2021-A09, CMS-CSST-2021-B06), the National Key Research and Development Program of China (22022YFA1603102), and the High-performance Computing Platform of Peking University.
This research was carried out in part at the Jet Propulsion Laboratory, which is operated by the California Institute of Technology under a contract with the National Aeronautics and Space Administration
(80NM0018D0004).
GCG acknowledges support by UNAM-PAPIIT IN103822 grant.
AP and EV-S acknowledge financial support from the UNAM-PAPIIT IG100223 grant. AP further acknowledges further support from the Sistema Nacional de Investigadores of CONACyT, and from CONACyT grant number 86372 of the `Ciencia de Frontera 2019’ program, entitled `Citlalc\'oatl: A multiscale study at the new frontier of the formation and early evolution of stars and planetary systems’, M\'exico.
MJ acknowledges support from the Academy of Finland grant No. 348342.
This work is supported by the international partnership program of Chinese Academy of Sciences through grant No.114231KYSB20200009, and Shanghai Pujiang Program 20PJ1415500.
G.G., AS and L.B. gratefully acknowledges support by the ANID BASAL projects ACE210002 and FB210003.
This paper makes use of the following ALMA data: ADS/JAO.ALMA\#2019.1.00685.S and 2015.1.01539.S. ALMA is a partnership of ESO (representing its member states), NSF (USA), 
and NINS (Japan), together with NRC (Canada), MOST and ASIAA (Taiwan), and KASI (Republic of Korea), in cooperation with the Republic of Chile. The Joint 
ALMA Observatory is operated by ESO, AUI/NRAO, and NAOJ.
This research made use of astrodendro, a Python package to compute dendrograms of Astronomical data ({\url{http://www.dendrograms.org/}}).
This research made use of Astropy,
a community-developed core Python package for Astronomy \citep{Ast18}. \par


\appendix
\section{Generating de-noised data cube of \htcop~(1--0), and moment maps} 
\label{app:denoise_moms}
To reduce the effect of the noise of the observed data on gas emission analysis, we produced a de-noised/modelled \htcop\ data cube using the BTS algorithm \citep{Cla18}.  It is an automated routine that can model multiple (up to six) velocity components in spectral lines. Here, the number 
of the components and their positions (in the velocity/frequency axis) are determined by deriving the 1st, 2nd and 3rd derivatives of a spectrum. The routine then adopts a least-squared fitting approach to refine the fit with the determined component number,  checking for over-fitting and over-lapping velocity centroids. 
As default inputs to BTS, we used $\alpha_{\rm BTS}=3$ (the channel number over which the spectrum is smoothed, see \citealt{Cla18}), $\beta_{\rm BTS}=5$ (signal-to-noise threshold for a velocity component) and
$\gamma_{\rm BTS}=1.5$ (fitting acceptance threshold).
As outputs, the BTS algorithm yields the amplitude ($I_{\rm c}$, in units of Jy~beam$^{-1}$~\vel), velocity centroid ($V_{\rm c}$ in units of \vel) 
and dispersion ($\sigma_{\rm c}$ in units of \vel) of each component, c. 
Based on the average spectrum of \htcop~(1--0) over the central hub region, the velocity range of the modelled spectra was set to be
[-57.5, -52.5]\,\vel. In this range, the fraction of the one and two velocity components in the total fitted spectra  is 96\%, and 5\%, respectively. This result indicates that \htcop~(1--0) line emission of the central hub region 
is characterized mostly as a single velocity component, which interprets well the observed single-peaked average spectrum over the same region (see above). 

Following \citet{Pan23}, the moment maps (Moment\,0, 1, and 2) were created from the modelled/de-noised \htcop~(1--0) cube that was constructed from the BTS-modelled parameters ($I_{\rm c}$, $V_{\rm c}$, and $\sigma_{\rm c}$).
The velocity-integrated intensity, Moment\,0, satisfies $m0 = \int I {\rm dV}$ where the integration is conducted over [-57.5, -52.5]\,\vel. Moment\,1,  the intensity-weighted mean velocity, was given by 
the form $\sum_{i=1}^{N}{m0_{\rm c,i}\ V_{\rm c, i}}/\sum_{i=1}^{N}{m0_{\rm c,i}}$ for $N$ components ($m0_{\rm c,i}$ and $V_{\rm c, i}$ referred as to the integrated intensity, and velocity centroid of the $ith$ component, respectively). With the similar weighting approach, Moment\,2, defined as the velocity dispersion, was calculated as $\sum_{i=1}^{N}{m0_{\rm c,i}\ \sigma_{\rm c, i}}/\sum_{i=1}^{N}{m0_{\rm c,i}}$ for 
$N$ components ($\sigma_{\rm c, i}$ corresponding to the velocity dispersion of the $ith$ component).

\section{Cloud-scale gas infalling/accretion rates} 
\label{app:accretion_rates}

Following \citet{Kir13}, the filament-rooted (parallel) mass inflow rate can be estimated by assuming 
a cylinder of filament mass $M_{\rm fil}$, and length $L_{\rm fil}$ at an inclination angle of $\phi$ to the plane of the sky (PoS) ($0\degr$ for being parallel to the PoS and $90\degr$ for being along the line of the sight (LoS), $0\degr$ used here). The filament-aligned (parallel) mass accretion rate, 
$\dot{M_{\parallel}}$, is written in the form:
$\dot{M_{\parallel}} = \frac{M_{\rm fil} \nabla V_{\rm \parallel, fil}}{\rm tan(\phi)},$
where $\nabla V_{\rm \parallel, obs}$ is the filament-aligned velocity gradient, related to $V_{\rm \parallel,fil}$ via the form $V_{\rm \parallel,fil}/L_{\rm fil}$.

The filament mass for F1--F3 was derived from the $18\arcsec$ H$_2$ column density ($N(\rm H_2)$) map that was created from {\it Herschel} data. The details of the map making are referred to \citealt{Per16,Pan23}. In a nut shell, 
the ratio of the {\it Herschel} 160\,\um\ to 250\,\um\
images was used as a dust temperature ($T_{\rm d}$) tracer and the derived temperature was then used together with the 250\,\um\ image to calculate
the column density ($N({\rm H_2})$) map, where the dust opacity law $\kappa_{\nu}=0.1 \times (\nu/1\,{\rm THz})^{\beta}$ was adopted along with 
$\beta=2$, and a gas-to-dust mass ratio of 100 \citep{Bec90}. The $N(\rm H_2)$ map enables 
the full spatial coverage of the G310 HFS cloud, allowing to count the 
full mass of the hub-composing filaments. The boundary of each filament is depicted roughly based on a column density level of $1.5\times10^{22}$\,\pcmsq, as indicated in the grey polygon in Fig.\,\ref{fig:overview}. In practice, a constant level of $1.0\times10^{22}$\,\pcmsq\ as background emission was subtracted from the $N(\rm H_2)$ map for the mass estimate. Accordingly, the enclosed mass of each filament was estimated to be $3.6\pm0.7\times 10^3$\,\msun\ for F1, and $1.4\pm 0.3\times 10^3$\,\msun\ for F2, and $0.8\pm 0.2\times 10^3$\,\msun\ for F3.
Given the  $\nabla V_{\rm \parallel, obs}$ values (see Sect.\,\ref{subsec:result:vg}, and Fig.\,\ref{fig:vel:grad}a),  
we obtain longitudinal mass inflow rates (see Table\,2) of $(3.6\pm1.5)\times10^{-4}$\,\msun~yr$^{-1}$ for F1, $(2.4\pm2.4)\times10^{-4}$\,\msun~yr$^{-1}$ for F2, and $(1.4\pm0.6)\times10^{-4}$\,\msun~yr$^{-1}$ for F3.


\end{document}